\documentclass[twocolumn]{jpsj2} 
\title{Evidence for Unconventional Superconductivity in Arsenic-Free Iron-Based Superconductor FeSe : A $^{77}$Se--NMR Study}

\author{Hisashi \textsc{Kotegawa}$^{1}$\thanks{E-mail address: kotegawa@crystal.kobe-u.ac.jp}, Satoru \textsc{Masaki}$^{1}$, Yoshiki \textsc{Awai}$^{1}$, Hideki \textsc{Tou}$^{1}$, Yoshikazu \textsc{Mizuguchi}$^{2}$, and Yoshihiko \textsc{Takano}$^{2}$}

\inst{$^{1}$Department of Physics, Kobe University, Kobe 658-8530\\
$^{2}$National Institute for Materials Science, 1-2-1, Sengen, Tsukuba, 305-0047, Japan\\
}

\abst{
We report the results of $^{77}$Se--nuclear magnetic resonance (NMR) in $\alpha$-FeSe, which exhibits a similar crystal structure to the LaFeAsO$_{1-x}$F$_x$ superconductor and shows superconductivity at 8 K. The nuclear-spin lattice relaxation rate $1/T_1$ shows $T^3$ behavior below the superconducting transition temperature $T_c$ without a coherence peak.  The $T_1T=$ const. behavior, indicative of the Fermi liquid state, can be seen in a wide temperature range above $T_c$. The superconductivity in $\alpha$-FeSe is also an unconventional one as well as LaFeAsO$_{1-x}$F$_x$ and related materials. The FeAs layer is not essential for the occurrence of the unconventional superconductivity.
}

\kword{FeSe, superconductivity, NMR}

\begin{document}
\maketitle

The recent rapid development in new superconductors containing Fe element is remarkable.
Immediately after the discovery of superconductivity at 26 K in LaFeAsO$_{1-x}$F$_x$ (ZrCuSiAs-type structure),\cite{Kamihara} a much higher $T_c$ was discovered in Nd- or Sm-substituted isostructural systems and also under pressure.\cite{Ren,Chen,Kito,Takahashi}
Moreover, the superconductivity in oxygen-free systems such as (Ba$_{1-x}$K$_x$)Fe$_2$As$_2$ (ThCr$_2$Si$_2$ type structure) or Li$_{1-x}$FeAs has been discovered.\cite{Rotter,Wang}
LaFeAsO$_{1-x}$F$_x$ and BaFe$_2$As$_2$ undergo magnetic ordering at low temperatures accompanied by the structural transition, implying that magnetic interactions are related with the occurrence of superconductivity.\cite{Cruz,Rotter2}
However, the mechanism of superconductivity in FeAs-based systems is a controversial issue.

Quite recently, Hsu {\it et al.} have discovered superconductivity at 8 K in $\alpha$-FeSe.\cite{Hsu}
This crystal structure belongs to the tetragonal symmetry  (PbO-type structure:$P4/nmm$) at room temperature.
Interestingly, $\alpha$-FeSe is composed of stacking layers of FeSe, similarly to LaFeAsO$_{1-x}$F$_x$ and related materials with FeAs layers.
The band structure calculated by Subedi {\it et al.} shows remarkable similarity to those of the FeAs-based superconductors.\cite{Subedi}
Furthermore, the $T_c$ in FeSe shows a rapid increase under pressure, reaching an onset value of 27 K at 1.48 GPa.\cite{Mizuguchi}
It is of great interest to investigate whether the superconductivity in $\alpha$-FeSe has the same origin as those in materials containing FeAs layers.

Here, we report the symmetry of superconducting gap and the magnetic properties in the normal state in $\alpha$-FeSe by $^{77}$Se--nuclear magnetic resonance (NMR) measurement.
A polycrystalline sample is prepared by the solid state reaction method, as described in refs.~12 and 13, and powdered for NMR measurement.
X-ray diffraction and neutron diffraction measurements using the sample made by the same procedure show that the actual composition of the sample is FeSe$_{0.92}$ due to a deficit of Se and that an impurity phase (hexagonal $\beta$-FeSe) exists .\cite{Mizuguchi,Margadonna}
Note that the sample reported by Hsu {\it et al.} was FeSe$_{0.88}$, but both our and Hsu {\it et al.}'s samples show zero resistance at $T_c=8$ K.
The resistivity and magnetization of our sample are reported in ref.~12.
The diamagnetic response of the sample is not large,\cite{Mizuguchi} but we show through NMR measurement that the superconductivity in FeSe is intrinsic.
The NMR measurement was performed by a standard spin-echo method.

\begin{figure}[b]
\centering
\includegraphics[width=0.75\linewidth]{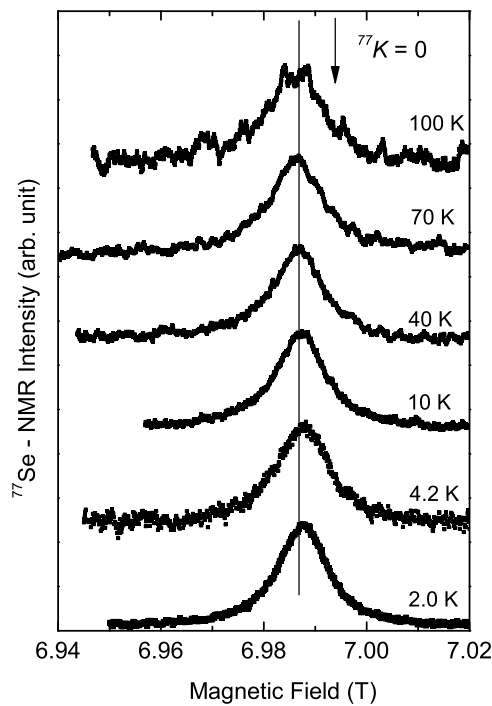}
\caption[]{
$^{77}$Se--NMR spectra measured at the NMR frequency of 56.86 MHz for FeSe. The arrow indicates $K=0$ for $^{77}$Se nucleus. The spectral shape is not a typical powder pattern but of a simple Lorentzian type. A small shift was observed below $T_c(H=7 {\rm T})\sim6$ K.
}
\end{figure}

Figure 1 shows the $^{77}$Se--NMR spectra measured at $\sim7$ T.
$^{77}$Se has a nuclear spin of $I=1/2$, a gyromagnetic ratio of $\gamma_N=8.13$ MHz/T, and a natural abundance of 7.5\%.
There is no significant difference in the spectra between 100 and 2 K.
The spectrum is composed of one Lorentzian shape, and shows no typical powder pattern with an anisotropic Knight shift, indicative of an isotropic Knight shift.
The Knight shift and linewidth are displayed in Fig.~2.
The linewidth is estimated by Lorentzian fitting.
The Knight shift shows a slight decrease with decreasing temperature, but is almost temperature-independent.
The Knight shift ($K$) is composed of the spin part ($K_s$) and temperature-independent orbital part ($K_{orb}$), that is,  $K(T)=K_s(T)+K_{orb}$.

\begin{figure}[htb]
\centering
\includegraphics[width=0.8\linewidth]{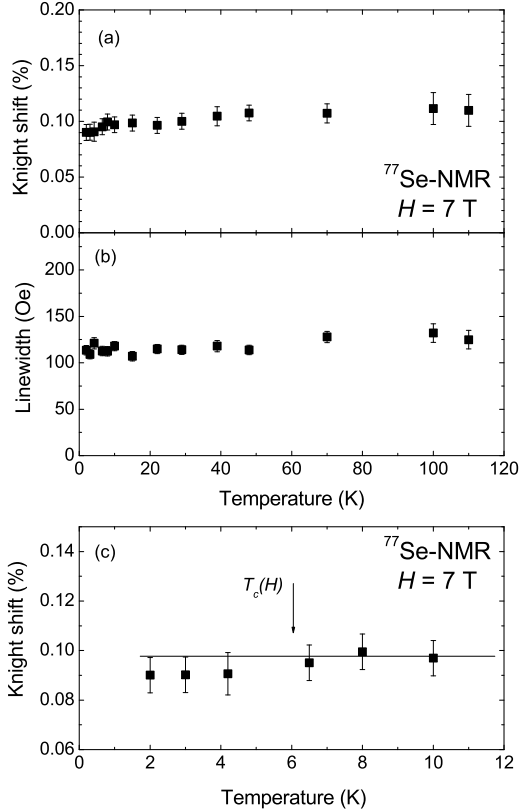}
\caption[]{
Temperature dependences of (a) Knight shift and (b) linewidth. There is no remarkable change in the Knight shift and linewidth at around the structural transition temperature $T_s=100$ K or 70 K. Linewidth is almost temperature-independent down to the lowest temperature, suggesting the absence of static magnetic ordering. (c) Knight shift below 12 K. The slight decrease in $K$ is observed below $T_c(H=7 {\rm T})\sim6$ K.
}
\end{figure}

In the normal state, the weak temperature dependence of $K$ suggests that $K$ is predominated by $K_{orb}$, or that temperature dependence of $K_s$ is weak.
Note that the static susceptibility is also almost independent of temperature.\cite{Mizuguchi}
However, because of the weak temperature dependence of $K$, it is difficult to distinguish between the spin part and orbital part in Knight shift from the present data.
The X-ray diffraction measurement shows a structural phase transition from tetragonal to orthorhombic at around $T_s=100$ K or 70 K.\cite{Hsu,Margadonna} 
At the present stage, we observe no corresponding significant anomaly.
The linewidth is also temperature-independent, as shown in Fig.~2(b).
It is noted that there is no signature of magnetic ordering below $T_s$, although antiferromagnetic ordering occurs in low-doping LaFeAsO$_{1-x}$F$_x$ and BaFe$_2$As$_2$ systems accompanied by structural transition.\cite{Cruz,Rotter2}

At low temperatures, the spectrum shows the slight shift to a higher field side in the superconducting state.
The shift between 10 and 2 K is estimated to be $\sim5$ Oe.
The linewidth is unchanged below $T_c(H)$ within an experimental error of 5 Oe, indicative of a large penetration depth $\lambda$.
If we assume that the broadening by diamagnetism is within experimental error, $\lambda$ is roughly estimated to be $3600\sim5000$ \AA.
Using the coherence length $\xi=35$ \AA~estimated from the upper critical field,\cite{Mizuguchi} we obtain a diamagnetic field of $\sim2$ Oe.
At the present stage, it is difficult to conclude whether the shift of 5 Oe originates from the decrease in $K_s$.
Although we decreased the magnetic field down to 2 T, the shift of the spectrum was small as well.
In PrFeAsO$_{0.89}$F$_{0.11}$, on the other hand, the decrease in Knight shift for the $^{75}$As nucleus was observed below $T_c$ owing to the spin-singlet pairing.\cite{Matano}
In the case of FeSe, Knight shift for the $^{77}$Se nucleus may be predominated by the orbital part.
Actually, band calculation suggests that the density of state at the Fermi level is predominated by the Fe $d$-orbit, and that the contributions of Se orbits are quite small.\cite{Harima}
Further careful measurement using a high-quality single crystal is needed for the determination of spin singlet or spin triplet.

\begin{figure}[htb]
\centering
\includegraphics[width=0.8\linewidth]{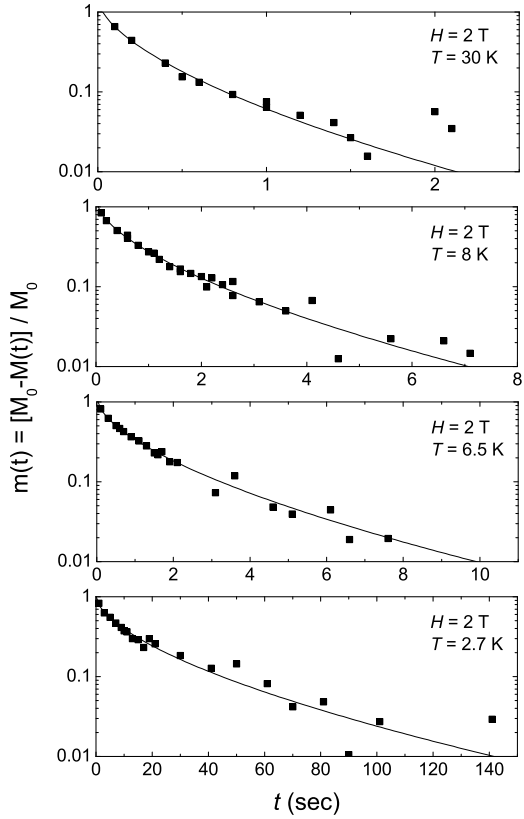}
\caption[]{
The recovery curves measured at 30, 8, 6.5, and 2.7 K under 2 T. The $m(t)=[M_0-M(t)]/M_0$ does not follow the single exponential function probably owing to an imperfection in the crystal structure, the contribution of the impurity phases, or the anisotropy of $1/T_1$. The solid curves are the fitting function of $m(t)=A\exp(-(t/T_1)^\alpha)$ with $\alpha=0.6$. This function follows $m(t)$ at all temperatures even below $T_c(H=2 {\rm T})=7$ K.
}
\end{figure}

The nuclear-spin lattice relaxation time $T_1$ was measured at the peak of the spectrum.
The recovery curves at several temperatures are shown in Fig.~3.
The $m(t)=[M_0-M(t)]/M_0$ does not follow a single exponential function expected in the case of the $-1/2 \iff 1/2$ transition for the $I=1/2$ nucleus.
We consider that this is attributed to an imperfection in the crystal structure, the contribution of impurity phases, or the anisotropy of $1/T_1$.
To determine $T_1$, we tentatively used the stretch type function $m(t)=A\exp(-(t/T_1)^\alpha)$.
We fixed the stretch coefficient $\alpha=0.6$, because the accuracy of our data did not allow us to treat $\alpha$ as a fitting parameter. 
As shown in the figure, this function follows $m(t)$ well in a wide temperature range even below $T_c$.
If the contribution from the impurity phases is significant, the fitting function changes markedly with temperature, particularly below $T_c$.
The unchanged fitting function ensures that the obtained $T_1$ mainly originates from the majority phase, that is, $\alpha$-FeSe.

\begin{figure}[htb]
\centering
\includegraphics[width=0.85\linewidth]{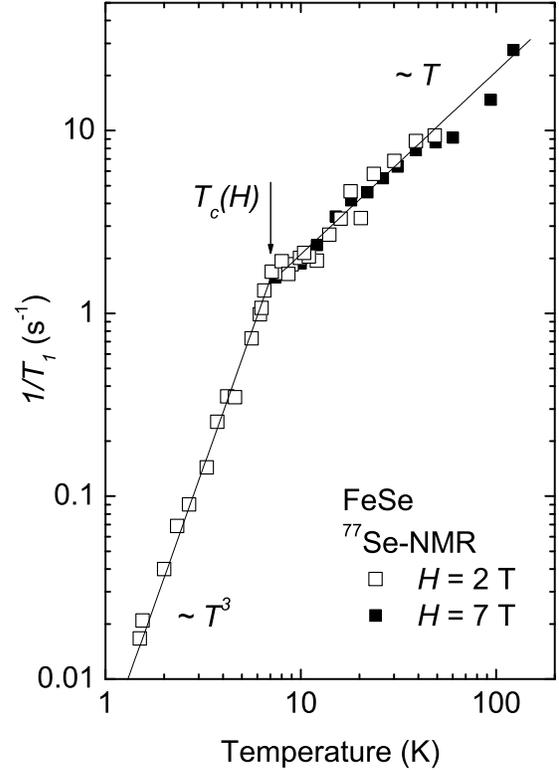}
\caption[]{
Temperature dependences of $1/T_1$ at 2 T and 7 T. The data at 7 T are plotted only above $T_c$. The $1/T_1$ exhibits the $T^3$ dependence below $T_c(H)$ and is proportional to $T$ above $T_c(H)$. There is no coherence peak just below $T_c(H)$.
}
\end{figure}

The nuclear-spin lattice relaxation rate $1/T_1$ measured at 2 T is displayed in Fig.~4.
The data at 7 T above $T_c$ are also plotted.
The $T_c$ at 2 T is confirmed to be 7 K by the $insitu$ ac susceptibility measurement using an NMR pick-up coil.
$1/T_1$ shows a drop just below $T_c$ and follows the $T^3$ dependence down to 1.5 K ($\sim0.2T_c$).
The fitting function for the recovery curve shows no distinct change even just below $T_c$ (6.5 K), as shown in Fig.~3.
There is no signature of a coherence peak that is characteristic of the phonon-mediated conventional $s$-wave superconductor.
These experimental facts suggest that the superconductivity in $\alpha$-FeSe is anisotropic with lines of vanishing gap on the Fermi surface.
The temperature dependence of $1/T_1$ below $T_c$ is quite similar to other NMR data of LaFeAsO$_{1-x}$F$_x$ and LaFeAsO$_{0.6}$.\cite{Nakai,Grafe,Mukuda}
The magnitude of superconducting gap $\Delta$ depends on the assumed gap function, but we confirmed that $2\Delta/k_BT_c$ in FeSe is almost the same as that in LaFeAsO$_{1-x}$F$_x$ ($x=0.11$).\cite{Nakai}
Within experimental accuracy, there is no structure indicating the multigap superconductivity below $T_c$ as seen in PrFeAsO$_{0.89}$F$_{0.11}$ with $T_c=45$ K.\cite{Matano}

In the normal state, $1/T_1$'s at 2 T and 7 T show almost the same temperature dependences.
The $1/T_1$ indicates the relation $T_1T=$ const. above $T_c$ up to at least $\sim50$ K.
The behavior of $T_1T=$ const. indicates the Fermi liquid state, in contrast with LaFeAsO$_{1-x}$F$_x$ and LaFeAsO$_{0.6}$.\cite{Nakai,Grafe,Mukuda}
In the case of LaFeAsO$_{1-x}$F$_x$ ($x=0.04$), $1/T_1T$ increases with decreasing temperature down to 30 K, showing a Curie-Weiss temperature dependence.\cite{Nakai}
In the case of LaFeAsO$_{1-x}$F$_x$ ($x=0.11$ and 0.10) and LaFeAsO$_{0.6}$, $1/T_1T$ decreases with decreasing temperature.\cite{Nakai,Grafe,Mukuda}
They have suggested the existence of a pseudogap.
The $1/T_1$ in FeSe shows no pseudogap behavior up to $\sim100$ K.

The unconventional superconductivity with $T_1T=$ const. has been reported in some materials such as Sr$_2$RuO$_4$, heavy-fermion systems CeIrIn$_5$ and CeCu$_2$Si$_2$ under pressures.\cite{Ishida,Kawasaki,Fujiwara} 
In these materials, it is controversial what triggers the unconventional superconductivity.
Our data for FeSe suggest that low-energy spin fluctuations are not essential for the occurrence of superconductivity.
To investigate why $T_c$ increases under pressure in FeSe is a clue to the elucidation of the superconducting mechanism.\cite{Mizuguchi}

In summary, we have investigated $^{77}$Se-NMR measurement in the newly discovered iron-based superconductor FeSe.
There is no signature of magnetic ordering down to the lowest temperature.
In the normal state, $1/T_1T$ is temperature-independent within experimental error, in contrast to the already-reported FeAs-based superconductors.\cite{Nakai,Grafe,Mukuda}
We found that FeSe is far from magnetic instability.
$1/T_1$ shows a rapid drop below $T_c$ without the coherence peak, following the $T^3$ behavior at low temperatures.
Our measurements suggest that the unconventional superconductivity with line nodes is realized in the Fermi liquid state in FeSe.
To our knowledge, this is the first binary compound to show unconventional superconductivity in $d$-electron systems.
We should note that recent theoretical works have pointed out the absence of the coherence peak and the power-law behavior in $1/T_1$ in the framework of a fully gapped $s_{\pm}$ scenario.\cite{Chubukov,Parker}
Our data clearly show the $T^3$ behavior below $T_c$, which is close to the line-node model, however, further careful investigations may be needed.
It is a future issue whether the superconductivity in FeSe is of completely the same origin as those in FeAs-based superconductors.
Nonetheless, this arsenic-free material opens a new route for the development of iron-based superconductivity.

We thank H. Harima for helpful discussions.
This work has been partly supported by Grant-in-Aids for Scientific Research (Nos. 19105006, 19204036, 19014016, 19051014, 19340103, 19014018, and 20045010) from the Ministry of Education, Culture, Sports, Science, and Technology (MEXT) of Japan. One of the authors (S.M.) has been financially supported as a JSPS Research Fellow.

\end{document}